# Dissipative particle dynamics simulation plus slip-springs for entangled polymers with various slip-spring densities

[a*]Yuichi Masubuchi, [bc]Hong-Xia Guo, [bc]Fan Wang,
[d]Bamim Khomami, [d]Mahdi Boudaghi-Khajehnobar,
[a]Yuya Doi, [a]Takato Ishida, and [a]Takashi Uneyama

[a] Department of Materials Physics, Nagoya University, Nagoya 4648603, Japan
[b] Beijing National Laboratory for Molecular Sciences, Joint Laboratory of Polymer Sciences and Materials, State Key Laboratory of Polymer Physics and Chemistry, Institute of Chemistry, Chinese Academy of Sciences, Beijing 100190, P. R. China;
[c] University of Chinese Academy of Sciences, Beijing 100049, P. R. China
[d] Department of Chemical & Biomolecular Engineering, University of Tennessee, Knoxville, Tennessee, 37996, USA

*to whom correspondence should be addressed
mas@mp.pse.nagoya-u.ac.jp
ver Jan 31, 2024

**Abstract**

Slip-spring models are valuable tools for simulating entangled polymers, bridging the gap between bead-spring models with excluded volume and network models with presumed reptation motion. This study focuses on the DPD-SS (Dissipative Particle Dynamics - Slip-Spring) model, which introduces slip-springs into the standard DPD polymer model with soft-core interactions. By systematically adjusting the fugacity of slip-springs, the density of slip-springs within the system is varied. Simulation results demonstrate the compatibility of models with different slip-spring densities in terms of diffusion and linear relaxation modulus when the average number of slip-springs per chain is the same. The conversion between DPD-SS models concerning length and time is achieved through Rouse scaling, which utilizes the average number of DPD beads between consecutive anchoring points of slip-springs. Additionally, the modulus conversion is accomplished through the plateau modulus that takes account of fluctuations around entanglement. Notably, diffusion and relaxation modulus from the DPD-SS model align with those reported for standard Kremer-Grest and DPD models featuring strong repulsive interactions.

**Introduction**

The dissipative particle dynamics (DPD) method[1–4] has been established as applicable to soft-matter systems. For the case of polymers[3], each polymer chain is represented by consecutive DPD beads connected by springs. The beads interact with each other according to the soft-core interaction[5]. The known problem of this standard DPD modeling is that the soft-core interaction does not prohibit crossing between polymers. Consequently, the entangled polymer dynamics are not reproduced. This issue was first reported by Spenley[6], who conducted a series of DPD simulations for polymer melts with the inter-particle interaction and connectivity for the chains proposed by Groot and Warren[3]. The results demonstrate that the molecular weight $N$ dependence of the longest relaxation time $\tau$ and the diffusion constant $D$ can be expressed as $\tau \propto N^{1.98 \pm 0.03}$ and $D \propto N^{-1.02 \pm 0.02}$ in $3 \leq N \leq 100$, without transition to entangled dynamics.

Pan and Manke[7] added the segmental repulsive potential (SRP) developed by Kumar and Larson[8] to the standard DPD interactions. They succeeded in reproducing the transition from the Rouse to the entangled behaviors; for $N > 30$, they reported $D \propto N^{-1.8}$ while retaining the Rouse behavior for the short chains. Goujon et al.[9] carefully examined this approach concerning chain crossability. Yamanoi et al.[10] utilized this method to report shear rheology. Sirk et al.[11] proposed a modified SRP by altering the inter-bond interaction for computational efficiency. Iwaoka et al.[12] considered an adjustable multipoint interaction to attain a cylindrical excluded bond volume.

The other direction was proposed by Nikunen et al.[13], who discussed the condition of uncrossability according to the relationship between the maximum bond stretch and the effective bead size defined by the intensity of repulsive force. Consequently, they realized the entangled polymer dynamics by intensifying repulsive interaction. This approach is straightforward and easily implemented in the conventional DPD method. However, the drawback is that the level of coarse-graining is close to that achieved by the modeling with Lenard-Jones-type hard-core interactions[14]. For instance, the critical molecular weight for the onset of entanglement in the molecular weight dependence of the diffusion coefficient is seen at $N \sim 50$, irrespective of the interaction parameter, according to Nikunen et al.[13] Groot[15] also examined a similar approach. Mohagheghi and Khomami[16] improved the model of Nikunen et al. by introducing bending potential along the chain to gain entanglement density and reduce entanglement molecular weight. They explored the entangled polymer dynamics under fast shear[16–18] and elongation[19] using this model (MK model, hereafter). Nafar Sefiddashti et al.[20,21] reported the careful mapping of the MK model with an atomistic model of polyethylene. The chain statistics and dynamics have recently been reported in further detail by Wang et al.[22] By adopting the same model, Li et al.[23] conducted a deep study on the slip phenomena and the individual configuration dynamics of interfacial copolymer chains in the entangled polymer－polymer interface under fast shear flow.

Another approach is the inclusion of slip springs. Likhtman[24] proposed a single-chain model in which a Rouse chain is trapped by several virtual springs representing dynamic constraints. For these additional springs, one end is anchored at a specific position in space, whereas the other end slides along the chain. Due to this sliding motion, the virtual spring is called a slip spring. Inspired by this single-chain model, multi-chain versions have been developed in parallel[25–30]. Among them, Langeloth et al.[28] installed slip springs into the DPD model with the standard inter-beads interaction (employed by Spenley[6]). This model reproduces the entangled polymer dynamics[28,31], even though the beads overlap. Utilizing the advantage of the DPD-based approach, they extended the model to polymer solutions, for which the polymers entangled via slip springs are mixed with DPD beads representing solvent[32]. Other studies[33,34] have also been reported with this model.

The distinguishable feature of the slip spring approach is that the level of coarse-graining can be chosen arbitrarily[35]. As established experimentally, the dynamics of different polymeric liquids are universal when the number of entanglements per chain is the same under equilibrium, irrespective of the number of Kuhn segments between entanglements[36]. This universality has been confirmed for a multi-chain slip-spring (MCSS) model[35]. Namely, the diffusion and the relaxation modulus are convertible when the number of anchoring points per chain is identical among the systems with different numbers of Rouse segments per chain. The conversion factors for length, time, and modulus[37] are theoretically derived according to the average number of Rouse beads between anchoring points. This feature allows us to adopt a suitable number of beads per chain according to the length and time ranges in focus and computation costs. However, the same is not guaranteed for other models that employ different inter-bead interactions like DPD interaction. Namely, the MCSS model is carefully designed to retain the Gaussian chain statistics by canceling the effect of slip springs via the specific soft-core inter-bead interaction.

In this study, we examined the DPD-based slip spring models. Hereafter, we refer to it as the DPD-SS model (meaning DPD with slip-springs). We confirmed that the arbitrariness for the coarse-graining level is observed, consistent with the MCSS model, despite the difference in the inter-bead interaction. If the conversion factors are chosen adequately, the attained polymer dynamics agree with those reported by the standard Kremer-Grest and the DPD simulations with strong repulsive interactions. Details are shown below.

**Model and Simulations**

We combine slip springs with the conventional DPD scheme[28,31,32]. Besides including slip springs, our model is essentially the same as the earlier DPD studies[2,3,6]. We consider polymer melts, in which each polymer chain is represented by consecutive DPD beads with the mass $m$. The force for particle $i$ is written as

$$\mathbf{F}_i = \mathbf{F}_i^C + \mathbf{F}_i^D + \mathbf{F}_i^R. \tag{1}$$

Here, $\mathbf{F}_i^C$, $\mathbf{F}_i^D$, and $\mathbf{F}_i^R$ are concervative, dissipative, and random forces, respectively.

$\mathbf{F}_i^C$ includes virtual force due to slip-springs in addition to conventional soft-core repulsive force and bonding force;

$$\mathbf{F}_i^C = \sum_{i \neq j} \mathbf{f}_{ij}^C + \sum_{\text{bonds}} \mathbf{f}_{ij}^B + \sum_{\text{slip-springs}} \mathbf{f}_{ij}^S. \tag{2}$$

$\mathbf{f}_{ij}^C$ is the soft-core repulsive interaction written as

$$\mathbf{f}_{ij}^C = \begin{cases} a_{ij}\left(1 - \dfrac{r_{ij}}{r_c}\right)\mathbf{e}_{ij} & (r_{ij} < r_c) \\ 0 & (r_{ij} \geq r_c) \end{cases}, \tag{3}$$

where $\mathbf{r}_{ij} = \mathbf{r}_i - \mathbf{r}_j$, $r_{ij} = |\mathbf{r}_{ij}|$, and $\mathbf{e}_{ij} = \mathbf{r}_{ij}/r_{ij}$. $a_{ij}$ is the repulsion parameter $a_{ij}$ chosen at $25k_BT/r_c$, and $r_c$ is the cut-off radius. $\mathbf{f}_{ij}^B$ is the bonding force between DPD beads connected along the chain written as

$$\mathbf{f}_{ij}^B = k\mathbf{r}_{ij}, \tag{4}$$

where $k$ is the spring constant chosen at $k = 2k_BT/r_c$.

$\mathbf{f}_{ij}^S$ is the contribution of slip-springs separately introduced from the bonding springs. For a pair of particles connected by a slip-spring,

$$\mathbf{f}_{ij}^S = k\mathbf{r}_{ij}, \tag{5}$$

where $k$ is the same spring constant as that for $\mathbf{f}_{ij}^B$. These slip-springs hop along the chain according to the cumulative probability below.

$$\Psi = \frac{k_B T \Delta t}{\zeta_s r_c^2}\left(1 - \tanh\frac{\Delta F}{2k_B T}\right) \tag{6}$$

Here, $\zeta_s$ is the friction coefficient of hopping, and $\Delta t$ is the step size of numerical integration. $\Delta F$ is the energy difference according to the change in spring length induced by the hopping written below.

$$\Delta F = \frac{1}{2}k(\mathbf{u}'^2 - \mathbf{u}^2) \tag{7}$$

Here, $\mathbf{u}$ and $\mathbf{u}'$ are the end-to-end vectors before and after the hopping. When a slip spring reaches the chain end by the hopping, the spring is annihilated with the following cumulative probability.

$$\Psi_- = \frac{k_B T \Delta t}{\zeta_s r_c^2} \tag{8}$$

Vise versa, we create a new spring to connect a randomly selected chain end and one of the surrounding beads. A creation attempt is accepted with the following probability.

$$\Psi_+ = \exp(-k\,\mathbf{u}^2/2k_\mathrm{B}T) \tag{9}$$

Here, $\mathbf{u}$ is the end-to-end vector of the newly created slip spring. We attempt the creation procedure with the number of trials below to fulfill the detailed balance.

$$K = 2N_\mathrm{ends}\frac{4\pi}{3}r_\mathrm{cs}^3\rho_\mathrm{b}\frac{\Delta t}{\zeta_s}\xi \tag{10}$$

Here, $N_\mathrm{ends}$ is the number of chain ends, $r_\mathrm{cs}$ is the cut-off distance for the creation (different from $r_\mathrm{c}$), and $\xi$ is the fugacity of slip springs. Since $K$ is not an integer in general, we take $\lfloor K \rfloor$ or $\lceil K \rceil$ in each time step stochastically to have $K$ as the time average. The parameters are chosen according to the previous studies as $k_\mathrm{B}T\Delta t/\zeta_s r_\mathrm{c}^2 = 0.06$ and $r_\mathrm{cs} = 3r_\mathrm{c}$. $\xi$ is systematically varied to control the slip spring density. Note that overlapping and crossing between slip-springs along the chain is allowed. If slip-spring degeneracy is disallowed, the chain dynamics slow down due to the asymmetric exclusion process of slip-springs[38].

$\mathbf{F}_i^D$ and $\mathbf{F}_i^R$ are dissipative and random forces written as

$$\mathbf{F}_i^D = \sum_{i\neq j} -\gamma w^D(r_{ij})(\mathbf{e}_{ij}\cdot\mathbf{v}_{ij})\mathbf{e}_{ij}, \tag{11}$$

$$\mathbf{F}_i^R = \sum_{i\neq j} \sigma w^R(r_{ij})\theta_{ij}\mathbf{e}_{ij}. \tag{12}$$

Here, $w^D(r)$ and $w^R(r)$ are the weighting functions given by

$$w^D(r) = [w^R(r)]^2 = \begin{cases}(1-r)^2 & (r<r_c)\\ 0 & (r\geq r_c)\end{cases}. \tag{13}$$

$\mathbf{v}_{ij}$ is the relative velocity. $\theta_{ij}$ is a Gaussian random value obeying $\langle\theta_{ij}\rangle = 0$ and $\langle\theta_{ij}(t)\theta_{kl}(t')\rangle = (\delta_{ik}\delta_{jl}+\delta_{il}\delta_{jk})\delta(t-t')$. $\gamma$ and $\sigma$ are chosen to satisfy the fluctuation-dissipation relation.

$$\sigma^2 = 2\gamma k_\mathrm{B}T. \tag{14}$$

The other parameters for simulations are set as follows. The bead density $\rho_\mathrm{b}$ is fixed at $3r_\mathrm{c}^{-3}$. The bead number per polymer chain $N$ varies from 7 to 150 depending on the slip spring density. The number of polymers is $M = 200$ or 300 to achieve sufficient simulation box dimensions compared to the polymer size. The integration timestep for the DPD dynamics is $\Delta t_\mathrm{DPD} = 0.06\tau$, where $\tau = \sqrt{r_\mathrm{c}^2 m/k_\mathrm{B}T}$. Eight independent simulation runs were performed for each condition. After equilibration, we acquired the data for a sufficiently long time, which is at least 10 times longer than the longest relaxation time of each system. We choose units of length, time, and energy as $r_\mathrm{c}$, $\tau$, and $k_\mathrm{B}T$, and quantities reported hereafter are normalized according to these standard DPD units.

**Results and Discussion**

Figure 1 shows the quantity $N_{eSS}$ defined below.

$$N_{eSS} \equiv \frac{2M_{SS}}{NM} \tag{14}$$

Here, $M_{SS}$ is the number of slip springs in the system. The factor of 2 in the numerator stands for the two anchoring points of each spring, and hence $N_{eSS}$ is the average number of DPD beads between two consecutive anchoring points of the slip springs along the chain. In Fig 1 (a), $N_{eSS}$ is plotted against the DPD bead number per chain $N$ for various $\xi$ values. $N_{eSS}$ slightly decreases with increasing $N$, and it converges to a steady value depending on $\xi$ for the large-$N$ limit. Note that for the case of the original MCSS model, $N_{eSS}$ weakly increases with increasing $\xi$ before reaching steady values. This qualitative difference between the models is due to the intra-chain correlation.[25] Nevertheless, the $N$-dependence is relatively weak, and we discuss the asymptotic value at the large-$N$ limit $N_{eSS}^{\infty}$, following the previous study.

Figure 1 (b) shows $N_{eSS}^{\infty}$ as a function of $\xi$. As seen in the previous studym $N_{eSS}^{\infty}$ is in inverse proportion to $\xi$. This behavior is consistent with what has been reported for the original MCSS model, for which the following relation has been reported with neglecting intra-chain correlation.

$$\frac{1}{N_{eSS}^{\infty}} \sim 2\xi \rho_b \left(\frac{\pi}{k_s}\right)^{3/2} \tag{15}$$

For the parameters chosen for this specific study, we have $N_{eSS}^{\infty} \sim 0.16 \xi^{-1}$. This relation is drawn by the broken line in Fig 1 (b), demonstrating a good agreement with the DPD-SS results shown by the symbol. This correspondence implies that the difference in the inter-bead interaction is not critical for $N_{eSS}^{\infty}$.

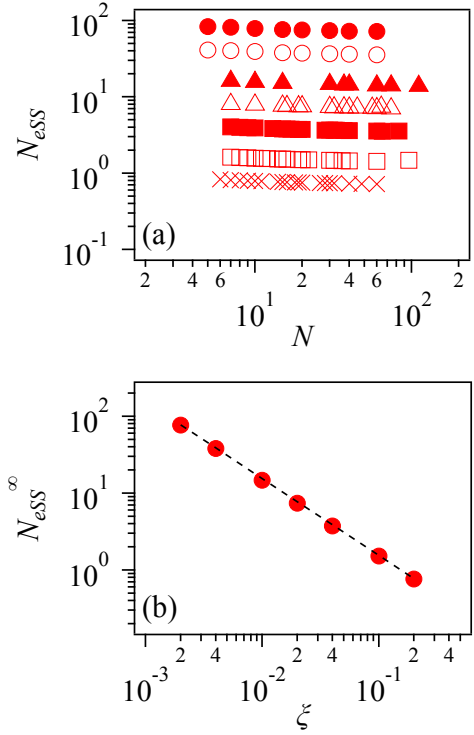

**Figure 1** Average number of DPD beads between consecutive anchoring points of slip springs $N_{eSS}$ plotted against the bead number per chain $N$ for various slip spring activities $\xi$ (a), and $N_{eSS}$ in the long chain limit $N_{eSS}^{\infty}$ as a function of $\xi$ (b). In Panel (a), $\xi = 0.002, 0.004, 0.01, 0.02, 0.04, 0.1,$ and $0.2$ from top to bottom. The broken line in Panel (b) shows the relationship $N_{eSS}^{\infty} = 0.16\xi^{-1}$.

Figure 2 shows the squared end-to-end distance $R^2$ normalized by the bond number $N-1$. For comparison, the case without slip springs ($\xi=0$) is also indicated by black unfilled circles and cross. (The latter indicates the data extracted from Spenley[6].) As clearly seen, DPD chains are not Gaussian with small $N$ values, and $R^2/(N-1) > 1$ even for the large-$N$ limit due to the repulsive interaction. This nature is also seen in DPD-SS results. With increasing $\xi$, the introduced slip springs compress the chains, and such an effect is seen for small-$N$ cases. Nevertheless, the change in $R^2$ statistics is relatively small compared to the distribution shown by the error bars for long chains with $N \geq 10$.

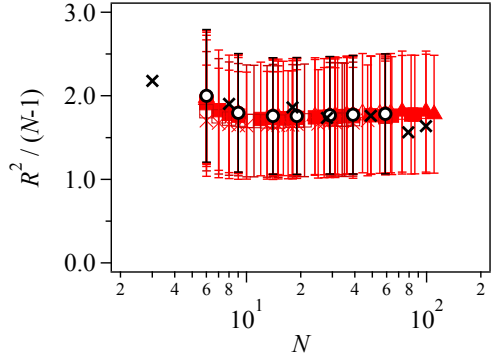

**Figure 2** The squared end-to-end distance $R^2$ normalized by the bond number per chain $N-1$ plotted against the bead number per chain $N$ with various slip spring activities $\xi$. The DPD-SS results for $\xi = 0.002, 0.004, 0.01, 0.02, 0.04, 0.1$, and $0.2$ are shown in red by filled circle, unfilled circle, filled triangle, unfilled triangle, filled square, unfilled square, and cross. The black unfilled circle shows the results for $\xi = 0$ (DPD without slip springs). The cross indicates the result reported by Spenley[6]. Error bars indicate the distribution.

Figure 3 (a) shows the diffusion coefficient $D$ as a function of $N$ for various $\xi$. To see the onset of entanglement, we plot $DN^2$ (the so-called Likhtman plot[24]), where the slopes of 1 and 0 correspond to the Rouse and reptation predictions. As reported earlier, $D$ follows the Rouse behavior when $\xi = 0$. As the value of $N$ increases, the entangled dynamics are realized by slip springs. As observed for polymers experimentally[39], the $N$-dependence of $D$ is more intense than the reptation theory[40] that predicts $D \propto N^{-2}$. Thus, $DN^2$ decreases with increasing $N$ in the large-$N$ range after showing a plateau. Irrespective of the value of $\xi$, the DPD-SS results are consistent with this established behavior. Figure 3 (b) demonstrates that the $DN^2$ curves with different $\xi$ values can be superposed when $D$ and $N$ are normalized by the values corresponding to the peak of $DN^2$. (Here, the results for $\xi \leq 0.004$ were omitted since they did not exhibit concave curves in the examined range of $N$.) Since the peak means the onset of entanglement, we refer to these characteristic values as $D_c$ and $N_c$.

The $N$−dependence of $D$ for DPD-SS is consistent with that reported for the standard bead-spring simulations and the earlier DPD simulations with strong repulsive interactions. In Fig 3 (b), these earlier results are also shown for comparison. The results from simulations of the bead-spring model proposed by Kremer and Grest (the KG model hereafter) are indicated by black symbols[14,41–44]. The plot is made with the critical values of $D$ and $N$ for the onset of entanglement for the KG model chosen at $D_{cKG} = 1.4 \times 10^{-4}$ (in units of Lenard-Jones liquids) and $N_{cKG} = 150$. Our results with various $\xi$ values are located within the scatter of those for the bead-spring simulations. The results from the DPD simulations reported by Nikunen et al.[13] with the repulsion parameter $a_{ij} = 100$ and the bead density $\rho = 1$ are shown by blue open circles, for which the critical values are assumed as $D_{cN} = 5.9 \times 10^{-4}$ and $N_{cN} = 110$. The blue and green open triangles are the data by Nafar Sefiddashti et al.[21] and Wang et al.[22] for the MK-DPD simulations with $a_{ij} = 200$, the bending potential parameter $k_b = 2$, and the bead density $\rho = 1$. For this case, the critical values are chosen as $D_{cMK} = 7.7 \times 10^{-4}$ and $N_{cMK} = 48$. These DPD results are also consistent with ours (in red) within a reasonable deviation range.

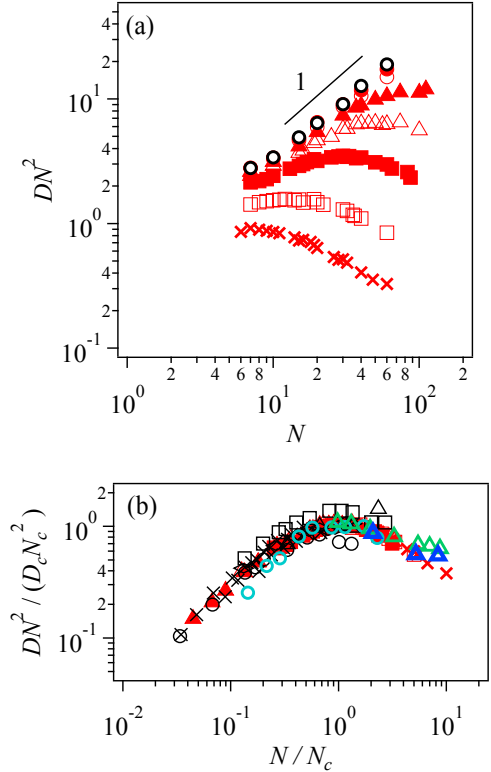

**Figure 3** Diffusion coefficient $D$ multiplied by the squared bead number per chain $N^2$ plotted against the bead number per chain $N$ for various slip spring activities $\xi$ (a), and those normalized by the characteristic values for the onset of entanglement $D_c$ and $N_c$ (b). The results for $\xi = 0.002$, 0.004, 0.01, 0.02, 0.04, 0.1, and 0.2 are shown by filled circle, unfilled circle, filled triangle, unfilled triangle, filled square, unfilled square, and cross in red. The black unfilled circle in Panel (a) shows the results for $\xi = 0$ (without slip springs). The solid line in Panel (a) indicates a slope of 1, showing the Rouse scaling. In Panel (b), the results for $\xi \leq 0.004$ are omitted. Black circles, crosses, triangles, squares, and rhombohedrals exhibit the results from literature for the Kremer-Grest model reported by Kremer and Grest[14], Bulacu and van der Giessen[41], Likhtman et al.[42], Xu et al.[43], and Takahashi et al.[44], respectively. The blue open circle, blue triangle, and green triangle show the DPD results by Nikunen et al.[13], Nafar Sefiddashti et al.[21] and Wang et al.[22], respectively.

Figures 4 (a) and (b) show $D_c$ and $N_c$ as functions of $\xi$. These panels exhibit the relations written as $D_c = 0.12\xi$ and $N_c = 1.1\xi^{-1}$. These relations can be converted to $D_c = 1.8 \times 10^{-2} N_{eSS}^{\infty}{}^{-1}$ and $N_c = 7.2 N_{eSS}^{\infty}$ according to eq 7. In Panel (c), we plot the relaxation time of the chain with the molecular weight of $N_c$, showing the relation written as $\tau_c = 0.35 N_c^2 = 0.42\xi^{-2}$, which is converted as $\tau_c = 18 N_{eSS}^{\infty}{}^2$. The results demonstrate that the chains with the molecular weight of $N_c$ exhibit the Rouse behavior. The conversion between the DPD-SS systems with different $\xi$ values can be achieved according to the Rouse scaling, as reported for the original MCSS model[35].

For the relation between $N_c$ and $N_{eSS}^{\infty}$, one may argue that for entangled polymers, the critical molecular weight $M_c$ is close to $2M_e$ and the relation shown here is inconsistent with this established, yet empirical, relation[36,40,45]. However, tabulated $M_e$ values are based on plateau moduli $G_N$, and the relationship between $G_N$ and $N_{eSS}^{\infty}$ is dependent on imposed fluctuations on the entanglement network[37,46–48].

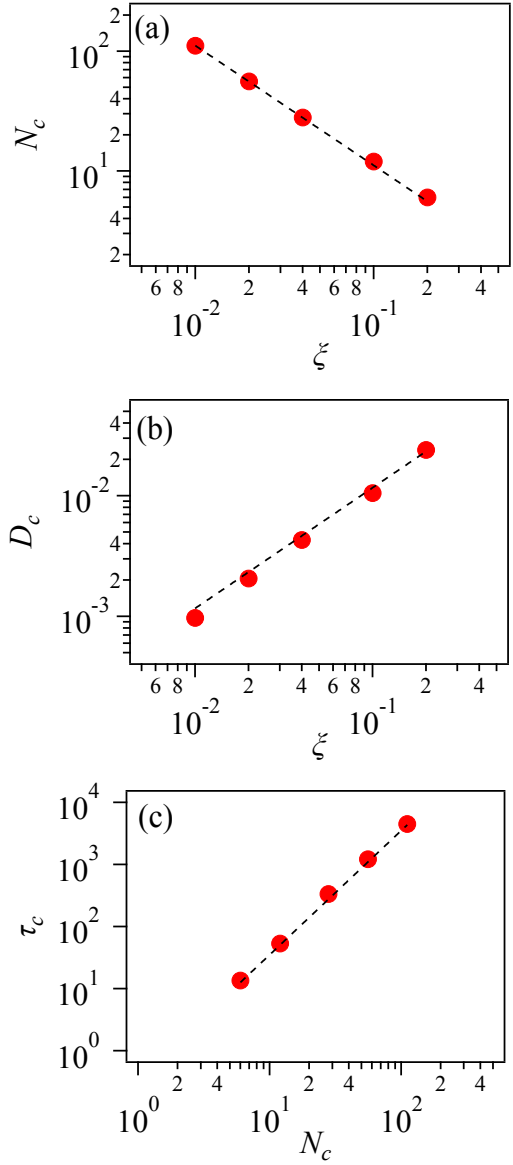

**Figure 4** The characteristic bead number per chain $N_c$ (a) and diffusion coefficient $D_c$ (b) for the onset of entanglement plotted against slip spring activities $\xi$. Panel (c) shows the end-to-end relaxation time $\tau_c$ for the chains with $N_c$ at each $\xi$. Black broken lines indicate $N_c = 1.1\xi^{-1}$, $D_c = 0.12\xi$, and $\tau_c = 0.35N_c^2$.

To further evaluate the compatibility among the systems with different $\xi$, we compare the mean-squared displacement of the bead located at the chain center $g_1(t)$. Figure 5 (a) shows reasonable agreement of DPD-SS results with some literature data[14,41,42,44,49] for the KG model with the bead number per chain of $N_{KG} = 50, 100, 200$, and $350$. As established earlier, $g_1(t)$ shows a few slope changes reflecting different subdiffusive motions before exhibiting Fickean diffusion. To see the transition from the diffusion along the chain contour ($g_1(t) \sim t^{1/2}$) to the normal diffusion ($g_1(t) \sim t$), we plot $g_1(t) t^{-1/2}$ in Figure 5 (b), following Likhtman[42]. DPD-SS simulations shown by solid curves nicely reproduce the data of the KG model indicated by symbols, irrespective of $\xi$ values. Here, the conversion is attained with the relations $(\sigma_{KG}/r_c)^2 = 13 N_{eSS}^{\infty}{}^{-1} = 84\xi$ and $\tau_{KG}/\tau = 1.6 \times 10^2 N_{eSS}^{\infty}{}^{-2} = 6.6 \times 10^4 \xi^2$, where $\sigma_{KG}$ and $\tau_{KG}$ are units of length and time for the KG model. These conversion rules for length and time are consistent with the diffusion coefficient. Namely, from the superposition shown in Fig 3 (b), we have $D_{cKG}/D_c = (1.4 \times 10^{-4})/(1.8 \times 10^{-2} N_{eSS}^{\infty}{}^{-1}) = 7.8 \times 10^{-3} N_{eSS}^{\infty} = 1.2 \times 10^{-3} \xi^{-1}$. This relation is close to what is obtained from the rules for length and time, $D_{KG}/D = (\sigma_{KG}/r_c)^2/(\tau_{KG}/\tau) = 8.1 \times 10^{-3} N_{eSS}^{\infty} = 1.3 \times 10^{-3} \xi^{-1}$. The conversion rule for length is also consistent with the conversion for molecular weight $N_{KG}/N = N_{cKG}/N_c = 21 N_{eSS}^{\infty}{}^{-1} = 1.3 \times 10^2 \xi$ if we take into account of the effect of repulsive interactions on the Kuhn length, as shown in Fig 2.

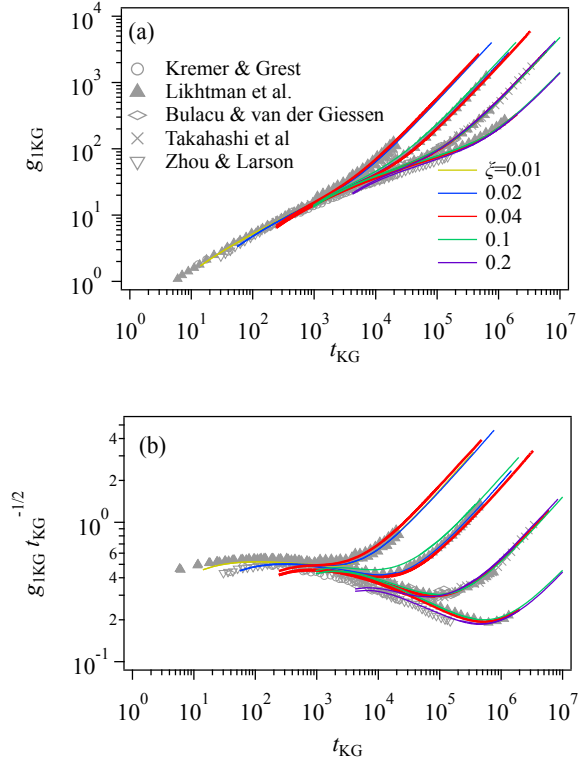

**Figure 5** Mean-squared displacement of the central bead $g_1(t)$ (a) and that normalized by Rouse behavior $g_1(t) t^{-1/2}$ (b) for various slip spring fugacities $\xi$ for different $N$ chains corresponding to Kremer-Grest chains with $N_{KG} = 50, 100, 200$, and $350$ from left to right. Symbols show the literature data for the KG model[14,41,42,44,49].

However, we note that the conversion factor for length determined via diffusion is inconsistent with the chain dimension. Figure 6 compares $R^2$ between DPD-SS and KG models. Although the $N-$dependence of $R^2$ is close to Gaussian in both cases, DPD-SS underestimates $R^2$ with the conversion factor $(\sigma_{\text{KG}}/r_c)^2 = 84\xi$. This discrepancy was also reported for the comparison of original MCSS model with the KG model[30,35], and the reason is the difference in chain density, chain stiffness, and inter-beads interactions[29,50].

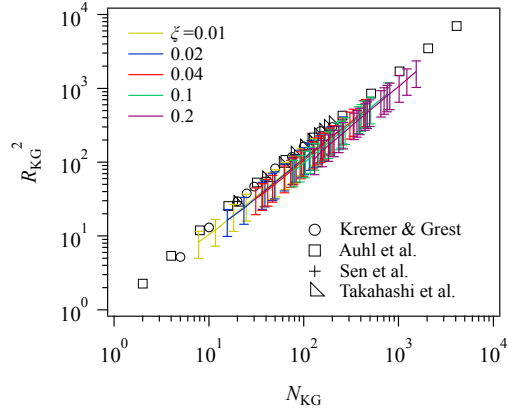

**Figure 6** End-to-end distance as a function of the bead number per chain for various slip spring fugacities $\xi$ converted to the Kremer-Grest model. Symbols show the literature data for the KG model[14,44,51,52]. Error bars indicate the distribution.

The other conversion factor we have to determine is that of modulus. Because modulus can be interpreted as thermal energy per unit volume, one may expect that the conversion of modulus is automatically achieved by the conversion factors for energy and length already determined. However, modulus depends on fluctuations in the entanglement network, and the conversion rule must be obtained separately. Figure 7 (a) shows linear relaxation modulus $G(t)$ for various $\xi$ values compared with the literature data[42,44] for melts of Kremer-Grest chains with $N_{\text{KG}} = 50, 100, 200,$ and 350. We omit the glassy part since such behavior is not included in the DPD-SS model. Consequently, the short-time behavior follows the Rouse prediction and $G(t) \sim t^{-1/2}$. After the Rouse behavior, long chains exhibit an entanglement plateau. For critical comparison, we plot $G(t)t^{1/2}$ in Figure 7 (b), in which convex curves reflecting entanglement can be seen for the long chains. DPD-SS simulations reproduce these established behaviors if a conversion factor for modulus is appropriately chosen for each $\xi$, as shown by curves. To be fair, we note that DPD-SS results for $N_{\text{KG}} = 200$ and 350 somewhat underestimate the data by Likhtman (triangle). However, the deviation is smaller than that observed between KG simulation data in literature. See the data by Takahashi et al.[44] (cross for $N_{\text{KG}} = 200$), for example.

According to the previous study[37,48], the plateau modulus of slip-spring systems is written as

$$G_N = \frac{11}{15} \frac{\rho}{N_{\text{eSS}}^\infty} \left\{ 1 + 4 \left( \frac{N_s}{N_{\text{eSS}}^\infty} \right) \right\}^{-\frac{1}{2}} \tag{16}$$

Due to this relationship between modulus and $N_{\text{eSS}}^\infty$, the conversion factor for modulus is rather complicated and not written as a power-law function of $N_{\text{eSS}}^\infty$. Instead, we have found that the conversion rule can be described as $G_{\text{KG}}/G = 2.6 \times 10^{-2} G_N^{-1}$. This relation is consistent with the previous study for the original MCSS model.[35]

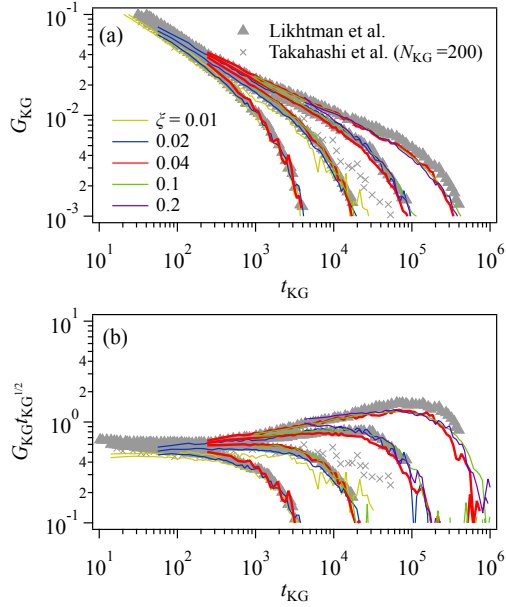

**Figure 7** Linear relaxation modulus $G(t)$ (a) and that normalized by Rouse behavior $G(t)t^{1/2}$ (b) for various slip spring activities $\xi$ for different $N$ chains corresponding to Kremer-Grest models with $N_{\text{KG}} = 50, 100, 200,$ and $350$ from left to right. Symbols show the literature data[42,44].

Let us turn our attention to the comparison with the entangled DPD model, in which the strong repulsive interaction realizes uncrossability between chains. We discuss the model proposed by Mohagheghi and Khomami[16], who introduced bending rigidity to reduce the entanglement molecular weight. Following the strategy we attained above for the KG model, we have determined the conversion factors as $(r_{c\text{MK}}/r_c)^2 = 53\xi$, $\tau_{\text{MK}}/\tau = 8.3 \times 10^3 \xi^2$, $N_{\text{MK}}/N = 42\xi$, and $G_{\text{MK}}/G = 1.2 \times 10^{-2} G_N$. Here, subscript MK refers to as the Mohagheghi-Khomami model. The DPD-SS results are shown only for $\xi = 0.2$ owing to the universality among different $\xi$ values.

The comparisons for $g_1(t)$ and $g_1(t)t^{-1/2}$ are shown in Fig. 8 for several cases with various $N_{\text{MK}}$. Most $g_1(t)$ data are from the literature, but a few are newly added for this study. The MSD from DPD-SS simulations (red curves) reasonably agrees with that from MK simulations for short chains with $N_{\text{MK}} \leq 150$, as seen in Fig 5 for the comparison to KG simulations. However, behavior for the longest chain with $N_{\text{MK}} = 400$ is different from that in MK. The reason for this discrepancy is unknown, and we need further evaluation for well-entangled cases.

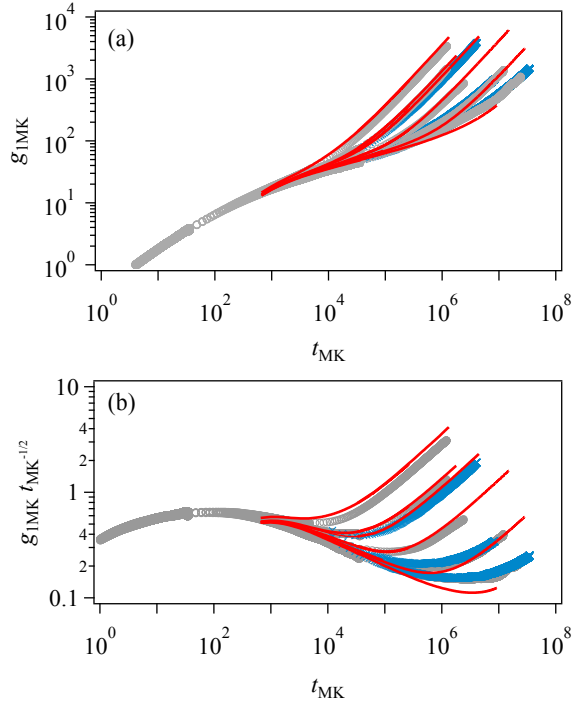

**Figure 8** Comparison between DPD-SS and MK simulations for the mean-square displacement of the central bead $g_1(t)$ (a) and that normalized by Rouse behavior $g_1(t)t^{-1/2}$ (b). The chain lengths were $N_{MK} = 60, 90, 100, 150, 200,$ and $400$, from left to right. The blue cross shows the data reported by Nafar Sefiddashti et al.[21] for $N_{MK}$=400, and newly added ones for $N_{MK}$=100 and 200. The gray circle indicates the data from Wang et al.[22]. Note that the MK results include two datasets for $N_{MK} = 200$ and 400. The red curves are the DPD-SS results with $\xi = 0.2$.

Figure 9 shows the comparisons for chain dimension and relaxation modulus. Panel (a) demonstrates that the conversion factor for length $(r_{cMK}/r_c)^2$ is not compatible with the chain dimension, as also seen in the comparison to the KG model (see Fig 6). As mentioned above, the reason for this discrepancy is the difference in chain density, chain stiffness, and inter-bead interactions[29,50]. Concerning $G(t)$, because the calculation is challenging for fine-grained models, correspondence between the models is not excellent. Nevertheless, a reasonable agreement is confirmed in Panel (b).

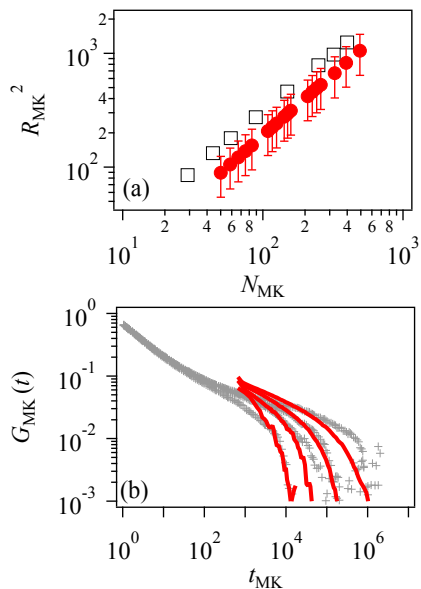

**Figure 9** Comparisons for the squared end-to-end distance $R^2$ (a) and the linear relaxation modulus $G(t)$ (b), between the MK simulations reported by Wang et al.[22] (black symbols) and our DPD-SS simulations with $\xi = 0.2$ (red curves and symbols). The bead number per chain for the MK model is $N_W =$ 60, 90, 150, and 250 in Panel (b). Error bars in Panel (a) indicate the distribution.

**Conclusions**

The effect of slip-spring fugacity on the polymer dynamics for the multi-chain slip-spring model with the standard DPD interaction (DPD-SS model) was examined. The slip-spring density depends on the fugacity as theoretically derived for the original MCSS model, implying that the difference in the inter-bead interaction does not have a significant effect on the thermodynamics of the model. For the dynamics, the onset of entanglement was nicely reproduced for the molecular weight dependence of the diffusion constant, irrespective of the fugacity, and the results obtained for different fugacities can be superposed with the conversion factors for the diffusion constant and molecular weight. The compatibility between the models with different fugacities was also observed for the mean-square displacement of the central bead in the polymer chain and linear relaxation modulus, with conversion factors for length, time, and modulus. The results are convertible to the literature data for the Kremer-Grest and Mohagheghi-Khomami simulations.

Although these results encourage applications of the DPD-SS model toward various problems, we note that the modeling has been established only for linear polymers with homogeneous chemistry in bulk under equilibrium. The chain dynamics under fast and large deformations[53,54] have yet to be discussed. Extensions toward branch polymers[55] and gelation[56] have been reported for the original MCSS but are not guaranteed for the model with the DPD interaction. Consecutive studies in such directions are ongoing, and the results will be reported elsewhere.

**Acknowledgments**

This study is partly supported by JST-CREST (JPMJCR1992) and JSPS KAKENHI (22H01189).